\documentclass[nofootinbib,preprint,
  showpacs,preprintnumbers,amsmath,amssymb]{revtex4}

\usepackage{latexsym}%
\usepackage{hyperref}%

\newcommand{\ben}{\begin{eqnarray*}}
\newcommand{\een}{\end{eqnarray*}}
\newcommand{\benn}{\begin{eqnarray}}
\newcommand{\eenn}{\end{eqnarray}}
\newcommand{\ba}{\begin{align}}
\newcommand{\ea}{\end{align}}

\newcommand{\arcsinh}{\mathrm{arcsinh}}

\begin{document}

\title{\bf Classical and quantum LTB model for the
 non-marginal case}
\author{Claus Kiefer}
\email{kiefer@thp.uni-koeln.de}
\affiliation{ Institut f\"ur Theoretische Physik,
Universit\"at zu K\"oln, Z\"ulpicher
Str. 77, 50937 K\"oln, Germany,\\
and Max-Planck-Institut f\"ur Gravitationsphysik,
Am M\"uhlenberg 1, 14476 Golm,
Germany.}
\author{Jakob M\"uller-Hill}
\email{jmh@thp.uni-koeln.de}
\affiliation{ Institut f\"ur Theoretische Physik, Universit\"at zu K\"oln,
Z\"ulpicher Str. 77, 50937 K\"oln, Germany.}
\author{Cenalo Vaz}
\email{vaz@physics.uc.edu}
\affiliation{Department of Physics, University of Cincinnati,
Cincinnati, Ohio 45221-0011, USA.}
\begin{abstract}
We extend the classical and quantum treatment of the
Lema\^{\i}tre--Tolman--Bondi (LTB) model to the
non-marginal case (defined by the fact that the shells of the
dust cloud start with a non-vanishing velocity at infinity).
We present the classical canonical formalism and address with
particular care the boundary terms in the action. We give the general
relation between dust time and Killing time.
Employing a lattice regularization, we then
derive and discuss for particular factor orderings exact solutions to all
quantum constraints.
\end{abstract}

\maketitle


\section{Introduction\label{intro}}

As long as a full quantum theory of gravity is not available,
it is important to address the quantization of particular models.
Among the most interesting ones are systems of spherically symmetric gravity
plus matter, because they are sufficiently non-trivial and
nevertheless simple enough for some technical treatment.
Moreover, they find physical applications in the study of
gravitational collapse and quantum aspects of black holes.

In this paper we discuss the Lema\^{\i}tre--Tolman--Bondi (LTB) model
which describes a self-gravitating dust cloud.
It has originally been introduced by Lema\^{\i}tre \cite{lemaitre}
in order to study cosmology where it has indeed found interesting
applications, cf. \cite{krasinski} for details and references.
Here the focus is on the canonical (Hamiltonian) formalism
for both the classical and the quantum LTB model.
The classical part is mainly intended as a preparation for the
quantum model but it exhibits interesting aspects onits own.

The physical questions that one eventually seeks to address
include the issue of singularity avoidance in the quantum theory,
the quantum evolution of black holes, and the role of the
naked classical singularities in quantum gravity.
In the simpler case of a single dust {\em shell}, for example,
it was shown by an explicit construction that a unitary quantum theory
exists in which the classical black-hole singularity is avoided,
cf. \cite{hajicek,OUP}. In fact, a wave packet describing a classical shell
evolves into a superposition of collapsing and expanding shell
and yields destructive interference at the place of the classical
singularity. The formalism of loop quantum gravity has also been applied
to spherically symmetric systems and it has been claimed
that the singularities are avoided \cite{loopqg}.

In recent years, earlier works on the canonical quantization
of the Schwarzschild black hole \cite{KT94,kuchar}
have been applied to develop a canonical description of
the collapse of a marginally bound, spherical,
time-like dust cloud \cite{VW99A,VWS01}. `Marginally bound' means
that the various shells of the dust cloud start with vanishing
velocity at infinity.
Quantization of this classical dust system leads to the
Wheeler--DeWitt equation
for the wave functional describing the quantum collapse. As special cases,
it was possible to obtain the Bekenstein mass spectrum and statistical entropy
of the charged and uncharged black hole \cite{VW01}.
Later, in \cite{VKSW03}, it was shown that the
semiclassical (WKB) treatment of the Schwarzschild black hole in this canonical
picture describes Hawking's thermal radiation and in \cite{VWS04}, going beyond
the WKB approximation, that quantum corrections
render the spectrum of the radiation non-thermal.

Here we again apply quantum geometrodynamics
and the Wheeler--DeWitt equation to the LTB
model in order to see how far this approach can be developed.
In contrast to the earlier works, we consider the generic case, that is,
the case including the non-marginal models for which the classical shells start
with a non-vanishing velocity at infinity.
This paper is organized
as follows: in Sec.~II we introduce the LTB model and present the
canonical formalism for the general models. Some
technical details are relegated to Appendix~A.
In Sec.~III we perform a lattice regularization and
find for a particular set of factor
orderings exact solutions to both the Wheeler--DeWitt equation
and the diffeomorphism constraint.
The uniqueness of this set of solutions
for the given ansatz is shown in Appendix~B.
Sec.~IV presents a brief
summary and an outlook on possible future work.


\section{The classical LTB model}

\subsection{Metric and classical solutions}

The LTB model describes a self-gravitating dust cloud.
Its energy-momentum tensor reads
$T_{\mu \nu} = \epsilon(\tau,\rho) u_{\mu} u_{\nu}$, where
$u^{\mu}=u^{\mu}(\tau, \rho)$ is the four-velocity vector
of a dust particle with proper time $\tau$ and labeled by $\rho$
($\rho$ thus labels the various shells that together form the
dust cloud).
The line element for the LTB spacetime is given by
\begin{align}
\label{ltb-metric}
\mathrm{d}s^2 &= -\mathrm{d}\tau^2 +
\frac{(\partial_{\rho}R)^2}{1+2E(\rho)} \mathrm{d}\rho^2
+ R^2(\rho)(\mathrm{d}\theta^2 + \sin^2\theta \mathrm{d}\phi^2)\ .
\end{align}
Inserting this expression into the Einstein equations leads to
\begin{align}\label{ltb-eg} 8\pi G\epsilon(\tau,\rho) =
\frac{\partial_{\rho}F}{R^2 \partial_{\rho}R}
  \quad \mathrm{and} \quad (\partial_{\tau}R)^2 = \frac{F}{R} + 2E
  \,,\end{align}
where $F(\rho)$ is some non-negative function,
$G$ is the gravitational constant, which we set equal to one
in the following (except for some key expressions where it is retained),
and we set $c=1$ throughout. The case of collapse
is described by $\partial_{\tau}R(\tau,\rho)<0$.

There exists still the freedom to rescale the shell index $\rho$.
This can be fixed by demanding
\begin{equation} \label{ltb-fix-r} R(0,\rho) = \rho \, , \end{equation}
so that for $\tau=0$ the label coordinate
$\rho$ is equal to the curvature radius $R$. Now we can express the functions
$F(\rho)$ and $E(\rho)$ in terms of the energy density $\epsilon$ at
$\tau=0$. From \eqref{ltb-eg} one gets
\begin{align} \label{ltb_F} F(\rho) &=8\pi
\int_0^{\rho} \epsilon(0,\tilde \rho)\,
\tilde{\rho}^2\,\mathrm{d}\tilde \rho\ , \\ E(\rho) &=\frac{1}{2}
[\partial_{\tau}R(\tau=0,\rho)]^2-\frac{4\pi}{\rho} \int_0^\rho
\epsilon (0,\tilde \rho) 
\tilde{\rho}^2 \, \mathrm{d}\tilde \rho \; .\end{align}
The interpretation of these quantities is that
$F(\rho)/2$ is the active gravitating mass inside of $R(\tau,\rho)$,
while $E(\rho)$ is the total energy of the shell labeled by $\rho$.
The marginally bound models are defined by $E(\rho)=0$.
In the present paper we discuss the general case which includes
the non-marginal case defined by $E(\rho)\neq0$.

The solution of \eqref{ltb-eg} is given by
\begin{align} \label{ltb-sol}
\tau-\tau_0(\rho) = - \frac{R^{\frac{3}{2}}(\tau,\rho)
Q\left(-\frac{2E(\rho) \, R(\tau,\rho)}{F(\rho)}\right)}{\sqrt{F(\rho)}} \,
,\end{align}
where $\tau_0$ is a constant of integration that
can be fixed by \eqref{ltb-fix-r} to read
\begin{align} \tau_0(\rho)=\frac{{\rho}^{\frac{3}{2}}Q(-\frac{2E
\, \rho}{F})}{\sqrt{F}} \, .\end{align}
The function $Q(y)$ is defined by the expressions
\begin{align} Q(y)&=
\left(\frac{\arcsin\left(\sqrt{y}\right)}{y^{\frac{3}{2}}}-\frac{\sqrt{1-y}}{y}\right)
\;\; \mathrm{for} \;\; 1 \geq y >0 \; ,\\ Q(y) &= \frac{2}{3} \;\;
\mathrm{for} \;\; y=0 \; ,\\
Q(y)&=\left(\frac{\arcsinh\left(\sqrt{-y}\right)}{(-y)^{\frac{3}{2}}}-\frac{\sqrt{1-y}}{y}\right)
\;\; \mathrm{for} \;\; 0>y\geq -\infty \; ,\end{align}
where $Q>0$ and $1\geq y \geq -\infty$.

Equation \eqref{ltb-sol} shows that at the dust proper time
$\tau=\tau_0(\rho)$ the shell labeled
by $\rho$ has reached a curvature radius $R=0$, that is, the physical
singularity. So $\tau$ can only take values between $-\infty$ and
$\tau_0(\rho)$.

\subsection{Hamiltonian formalism}

For the canonical formalism one starts with
the general ansatz for a spherically-symmetric line element,
\begin{align}\mathrm{d}s^2
  = -N^2 \mathrm{d}t^2 + L^2 \left( \mathrm{d}r + N^r \mathrm{d}t \right)^2 +
  R^2 \mathrm{d} \Omega^2  \; , \label{spheric_metric}\end{align}
where $N$ and $N^r$ are the lapse and shift function, respectively.
The canonical momenta are given by
\begin{align}\label{P_L} P_L &=
  \frac{R}{N} \left(- \dot{R} + N^r R' \right)\ , \\ P_R &= \frac{1}{N} \left[ -L
    \dot{R} - \dot{L}R + \left(N^r L R \right)'\right] \; . \label{P_R}\end{align}
A dot denotes a derivative with respect to coordinate time $t$,
while a prime denotes a derivative
with respect to $r$. All variables are functions of $t$ and $r$.

A Legendre transformation from the Einstein--Hilbert action leads to
\begin{align} S_{\mathrm{EH}} = \int
  \mathrm{d}t \int_0^{\infty} \mathrm{d}r \left(P_L \dot{L} + P_R \dot{R} - N
    H^g - N^r H_r^g \right) + S_{\partial \Sigma} \; ,\label{mega_action}\end{align}
where the Hamiltonian and the diffeomorphism (momentum) constraint are given by
\begin{align} H^g &= - G\left(\frac{P_L P_R}{R} - \frac{LP_L^2}{2 R^2}\right) +
  \frac{1}{G}\left[ -\frac{L}{2} - \frac{R'^2}{2L}+ \left(\frac{RR'}{L} \right)' \right]\ ,
  \\ H_r^g &= R' P_R - LP_L' \; ,\end{align}
respectively, and the boundary action $S_{\partial \Sigma}$ is discussed below.

The total action is the sum of \eqref{mega_action} and an action $S^d$
describing the dust. The canonical formalism for the latter
was developed in \cite{KUC}, cf. also \cite{VWS01}. It reads
\begin{align} S^d
  &= \int \mathrm{d}t \int_0^{\infty} \mathrm{d}r \left( P_{\tau} \dot{\tau} - N H^d -
    N^r H_r^d \right) \; ,\\ \intertext{where the Hamiltonian and momentum
    constraints are} \label{H_dust}H^d &= P_{\tau}
  \sqrt{1+\frac{{\tau'}^2}{L^2}} \quad \mathrm{and} \quad H_r^d = \tau'
  P_{\tau} \; . \end{align}
In principle one would prefer to take a fundamental field
(e.g. a scalar field) for the matter part. However, this would make
the formalism much less tractable \cite{Romano}. Moreover, the
main features discussed here already exhibit themselves for the
dust model.

\subsection{Mass function in terms of canonical variables}

In the following we shall write the mass function $F(\rho)$, which was
introduced in \eqref{ltb_F}, from the canonical data.
This is essential for deriving consistent falloff
conditions that are appropriate for a realistic collapse model.
For the Schwarzschild case this was done in \cite{kuchar},
while for the marginal solutions ($E=0$) this was done in
\cite{VWS01}. Here we shall extend the derivation to the non-marginal case.

We start by demanding that the spacetime described by the metric
\eqref{spheric_metric} be embedded in a LTB spacetime. Considering the
LTB metric \eqref{ltb-metric}, a foliation described by functions
$\tau(r,t)$ and $\rho(r,t)$ leads to
\begin{align} \mathrm{d}s^2 &=
  \begin{aligned}[t] - &\left( \dot{\tau}^2 -
  \frac{\dot{\rho}^2(\partial_{\rho}R)^2}{1+2E} \,\right) \mathrm{d}t^2 - \left(2
  \dot{\tau} \tau' - \frac{2 \dot{\rho}\rho'(\partial_{\rho}R)^2}{1+2E}
  \,\right) \mathrm{d}t \mathrm{d}r \\ & +
  \left(\frac{{\rho'}^2 (\partial_{\rho}R)^2}{1+2E}  -
  {\tau'}^2\right)\mathrm{d}r^2  + R^2 \mathrm{d}\Omega
  \; .\end{aligned}\end{align}
Comparison with \eqref{spheric_metric} gives the following three equations:
\begin{align} N^2 - L^2 {N^r}^2 &= \dot{\tau}^2 - \frac{\dot{\rho}^2 (\partial_{\rho}R)^2}{1+2E}
  \; , \label{lost_N_Nr}\\ L^2 N^r &=
  \frac{\dot{\rho}\rho'(\partial_{\rho}R)^2 }{1+2E} \, - \dot{\tau} \tau' \; ,
  \label{lost_Nr}\\ L^2 &= \frac{{\rho'}^2(\partial_{\rho}R)^2}{1+2E} -
  {\tau'}^2 \; .\label{lost_L} \end{align}
Eliminating $N^r$ from \eqref{lost_N_Nr} with the help of \eqref{lost_Nr}
and \eqref{lost_L} gives
\begin{equation} N^2 =
   (\partial_{\rho}R)^2 \,\frac{(\rho' \dot{\tau}-
  \tau' \dot{\rho})^2}{L^2(1+2E)} \; . \label{lapse_n}\end{equation}
Defining $\mathcal{F}\equiv 1-F/R$ we can rewrite Einstein's equation \eqref{ltb-eg} as
\begin{equation}
\label{friedmann}
  (\partial_{\tau}R)^2 = 2E + 1 - \mathcal{F} \;
  .\end{equation}
Now we insert the expressions for lapse function \eqref{lapse_n} and shift
  vector \eqref{lost_Nr} into the expression for the canonical momentum $P_L$ \eqref{P_L}. Defining $\bar{R} \equiv
  (\partial_{\rho}R)/\sqrt{1+2E}$ and $J \equiv
  \rho'\dot{\tau}-\dot{\rho}\tau'$ we have 
\begin{align}\frac{LP_L}{R} &= \frac{1}{\bar{R}J}\left(-\dot{R}L^2+
R'[-\dot{\tau}\tau'+\dot{\rho}\rho'\bar{R}^2]\right) \;
  .\end{align}
Using the expression \eqref{lost_L} for $L$ on the right-hand side as well as
\begin{align}\dot{R} &= \dot{\tau}\partial_{\tau}R
  +\dot{\rho} \partial_{\rho}R = \dot{\tau} \partial_{\tau} R+
  \bar{R} \dot{\rho} \sqrt{1+2E} \; ,\\ R' &= \tau' \partial_{\tau}R + \rho' \partial_{\rho}R = \tau'
  \partial_{\tau}R+\bar{R}\rho' \sqrt{1+2E} \; ,\end{align}
we get
\begin{align}\frac{LP_L}{R} \bar{R} J = -\bar{R}^2 J
  \rho' \partial_{\tau}R - \bar{R} 
  \sqrt{1+2E}\tau' J \; ,\end{align}
and therefore
\begin{align} \frac{LP_L}{R}
  &=-\frac{(R' - \tau'\partial_{\tau}R)\partial_{\tau}R}{\sqrt{1+2E}} -
  \sqrt{1+2E} \tau' \\ &\stackrel{\makebox[0cm]{\scriptsize \eqref{ltb-eg}}}{=}
  -\frac{R'\sqrt{2E+1-\mathcal{F}}}{\sqrt{1+2E}} -
  \frac{\mathcal{F}\tau'}{\sqrt{1+2E}} \; .\end{align}
Solving this equation for $\tau'$ gives
\begin{equation} \label{lost_tau}\tau' = -\frac{1}{\mathcal{F}} \left(R'
  \sqrt{1+2E-\mathcal{F}} + \frac{LP_L\sqrt{1+2E}}{R} \right) \;
  .\end{equation}
Inserting this expression into \eqref{lost_L} yields
\begin{align} L^2 &= \bar{R}^2 {\rho'}^2 -
  {\tau'}^2 \\ &=
  \frac{R'^2}{1+2E}-\frac{2R'\tau'}{1+2E}\sqrt{1+2E-\mathcal{F}}
  -\frac{\mathcal{F}\tau'^2}{1+2E}  \\ &=
  \frac{1}{\mathcal{F}}\left(R'^2- \frac{L^2 P_L^2}{R^2}\right) \; ,\end{align}
which leads to 
\begin{equation}\mathcal{F} =
  \frac{{R'}^2}{L^2} - \frac{P_L^2}{R^2} \; .\end{equation}
We can thus express $F$ locally in terms of the canonical data
as follows:
\begin{equation}\label{lost_F} F = R \left[ 1 + \frac{P_L^2}{R^2}-
  \frac{{R'}^2}{L^2} \right] \; .\end{equation}
This is the same expression as was obtained in
\cite{VWS01} for the marginal models. It thus possesses a much wider
range of applicability and holds, in fact, for {\em all} cases.

Since $R=F$ at the horizon, $\mathcal{F}=0$ there.
We can check that though $\mathcal{F}$ appears in the denominator of
\eqref{lost_tau}, $\tau'$ is well behaved at the horizon, as it should be:
\begin{align} \tau'
  &\stackrel{\mathcal{F}\rightarrow 0}{\longrightarrow} \frac{1}{2} \left(R' +
  L \right) \; .\end{align}
As in the case of the Schwarzschild black hole \cite{kuchar} and
the marginal LTB-model \cite{VWS01}, one can make a canonical
transformation in order to elevate the mass function $F$ to a canonical
coordinate. The expressions are the same as in these earlier papers.
The canonical transformation is
\begin{equation}(\tau, R, L, P_{\tau}, P_R, P_L) \longrightarrow (\tau, R,
  F, P_{\tau}, \bar{P}_R, P_F)\ , \end{equation}
where
\begin{equation}\bar{P}_R = P_R -
  \frac{LP_L}{2R}-\frac{LP_L}{2R\mathcal{F}} - \frac{\Delta}{RL^2\mathcal{F}}
  \; , \end{equation}
with
\begin{equation} \Delta =
  (RR')(LP_L)'-(RR')'(LP_L) \; .\end{equation}
The action in the new canonical variables then reads
\begin{align}S_{\mathrm{EH}} = \int \mathrm{d}t \int_0^{\infty} \mathrm{d}r
  \left(P_{\tau} \dot{\tau} + \bar{P}_R \dot{R} + P_F \dot{F} - N H - N^r H_r
  \right) + S_{\partial \Sigma} \, ,\end{align}
where the new constraints are
\begin{align}\label{H_const_new} H &= - \frac{1}{2L} \left(\frac{F'R'}{G \,\mathcal{F}}
    + 4 G \mathcal{F} P_F \bar{P}_R \right) + P_{\tau}
  \sqrt{1+\frac{{\tau}'^2}{L^2}} \; ,\\ H_r &= \tau' P_{\tau} + R' \bar{P}_R +F' P_F
  \; . \end{align}
We shall now discuss the boundary action $S_{\partial \Sigma}$
in more detail.

\subsection{Boundary action}

Boundary terms are obtained from a careful discussion of
the falloff conditions for the canonical variables.
This was investigated for the marginal case in \cite{VWS01}.
For the non-marginal case, this derivation remains unchanged
if $E\to 0$ for $r\to 0$.
But this follows if one assumes that the initial density profile
is regular at the center: Taking $R(0,r)=r$ (this is just a choice
of scaling), then
\begin{equation}
F(r)=\int r^2\mu(r)\ \mathrm{d}r\ ,
\end{equation}
where $\mu(r)=\epsilon(0,r)$ is some initial density profile, cf.
\eqref{ltb_F}.
For a profile that is regular at the center, choose
\begin{equation}
\mu(r)=\sum_{n=0}^{\infty}\mu_nr^n\ ,
\end{equation}
which gives
\begin{equation}
F(r)=\sum_{n=0}^{\infty}F_nr^{n+3}\ .
\end{equation} 
As the center of the collapsing cloud is taken to be at rest 
in a spherically symmetric collapse, the velocity profile
 $\partial_{\tau}R(0,r)$, 
must, to leading order, behave as some positive power of the 
label coordinate, $r$. 
It then follows from the `velocity equation' \eqref{friedmann}
that $E$ behaves as $r^2$ or higher. In particular, $E\to 0$ as $r\to 0$. 

The only boundary term is obtained from the variation with
respect to $L$ and reads
\begin{align} \int{\mathrm d}t \ N_+(t) \delta M_+(t) \; ,  \end{align}
where $N_+(t) \equiv N(t,r\to\infty)$ is the lapse function at infinity and
$M_+(t)\equiv F(r\to\infty)/2$ is the ADM mass. To avoid the conclusion
that $N_+(t)$ is constraint to vanish, which would freeze the evolution at
infinity, the boundary term has to be canceled by an appropriate boundary
action.  This can be achieved by adding the surface action
\begin{align} S_{\partial \Sigma} &= - \int \mathrm{d}t\
  N_+(t) M_+(t) \; .\end{align}
Since varying $N_+$ would lead to zero mass, Kucha\v{r} has argued in
\cite{kuchar} that $N_+$ has to be treated as a prescribed function.  The lapse
function gives the ratio of proper time to coordinate time in the direction
normal to the foliation. Since $N^r(r)$ vanishes for $r \rightarrow \infty$,
the time evolution at infinity is generated along the world lines of observers
with $r=\mathrm{const}$. If we introduce the proper time, $\bar{\tau}_+$,
of these observers as a new variable, we can express the lapse
function in the form $N_+(t) = \dot{\bar{\tau}}_+(t)$. This leads to
\begin{equation} S_{\partial
    \Sigma} = - \int \mathrm{d}t \, M_+ \dot{\bar{\tau}}_+ \; . \label{tau_bar_bound}\end{equation}
Thus we have removed the necessity of fixing the lapse function at
infinity. (In \cite{kuchar} this is called
`parametrization at infinities'.)

In \cite{VWS01}, the proper time $\bar{\tau}_+$ was identified with
the dust proper time at infinity, $\tau_+$. By realizing that $\bar{\tau}_+$
is a priori not directly related to $\tau_+$, we propose here a different
treatment. Extending the treatment from the Schwarzschild case
\cite{kuchar} to here, the aim is to cast the homogeneous part of
the action into Liouvilee form and to find a transformation to
new canonical variables that absorb the boundary terms. This
can be done by introducing the mass density $\Gamma \equiv F'$
as a new canonical
variable and using that $F(0)=0$ (which is appropriate for a
collapse situation).
Part of the  Liouville form can then be rewritten as follows:
\begin{align} \bar{\theta} &\equiv
\int_0^{\infty} \mathrm{d}r \, P_{F} \delta F - M_+
  \delta \bar{\tau}_+ \\ &= \int_0^{\infty} \mathrm{d}r \, P_F(r) \int_0^{r}
  \mathrm{d}r' \, \delta \Gamma(r')+ \bar{\tau}_+ \delta M_+  -  \delta (M_+ \bar{\tau}_+) \\
  &= \int_0^{\infty} \mathrm{d}r P_F(r) \int_0^{r} \mathrm{d}r' \delta
  \Gamma(r')+ \frac{\bar{\tau}_+}{2} \int_0^{\infty}\mathrm{d}r' \, \delta \Gamma(r') - \delta
  (M_+ \bar{\tau}_+)\\ &= \int_0^{\infty} \mathrm{d}r \, \delta \Gamma(r) \int
  _r^{\infty} \mathrm{d}r' \, P_F(r') + \int_0^{\infty} \mathrm{d}r' \delta
  \Gamma(r') \frac{\bar{\tau}_+}{2} - \delta (M_+ \bar{\tau}_+)\\ &=\int_0^{\infty} \mathrm{d}r \,
  \delta \Gamma \left( \frac{\bar{\tau}_+}{2} + \int_r^{\infty} \mathrm{d}r' P_F(r') \right) -
  \delta (M_+ \bar{\tau}_+) \label{new_liouville}\; .\end{align}
In the calculation we have used the identity
\begin{equation}\int_0^{\infty} \mathrm{d}r \, P_F(r)
  \int_0^r \mathrm{d}r' \, \delta \Gamma(r') = \int_0^{\infty} \mathrm{d}r \, \delta
  \Gamma(r) \int_r^{\infty} \mathrm{d}r' \, P_F(r') \; ,\end{equation}
which can be seen by integrating
\begin{equation}\left( \int_{\infty}^{r}
    \mathrm{d}r' P_F(r') \int_0^r \mathrm{d}r' \delta\Gamma(r')\right)' =
  P_F(r) \int_0^{r} \mathrm{d}r' \delta \Gamma(r') - \delta \Gamma(r)
  \int_r^{\infty} \mathrm{d}r' P_F(r') \end{equation}
from $0$ to $\infty$. The left-hand side gives zero and the rest the desired
equation.

 From line \eqref{new_liouville} we can read off that
$P_{\Gamma}=\bar{\tau}_+/2 + \int_r^{\infty} \mathrm{d}r P_F$.
Thus we see that
$P_{\Gamma}(\infty)=\bar{\tau}_+/2$.  As will become clear from the
next subsection, this means that the Killing time at infinity
matches the prescribed function $\bar{\tau}_+$. Thus the new action reads
\begin{align}
  \label{action_2} S_{\mathrm{EH}} &= \int \mathrm{d}t \int_0^{\infty} \mathrm{d}r
  \left(P_{\tau}\dot{\tau} + \bar{P}_R \dot{R} + P_{\Gamma} \dot{\Gamma} - N
    H^g - N^r H_r^g \right) \; .\end{align}
The constraints in the new variables are
\begin{align} \label{H1} H &= - \frac{1}{2L} \left(\frac{\Gamma R'}{G\mathcal{F}} - 4G \mathcal{F}P_{\Gamma}'
    \bar{P}_R \right) + P_{\tau} \sqrt{1+\frac{{\tau}'^2}{L^2}} \; ,\\ \label{H2} H_r &= \tau' P_{\tau} + R' \bar{P}_R
  -\Gamma P_{\Gamma}' \; .\end{align}
The Hamiltonian constraint can be greatly simplified if the momentum constraint
is used to eliminate $P_F\equiv -P_{\Gamma}'$. The details of the
calculation can be found 
in Appendix~A. The constraints \eqref{H1} and \eqref{H2} can then be
replaced by the following equivalent set,
\begin{align}\label{H_new} H &=
  G\left({P_{\tau}}^2 + \mathcal{F} \bar{P}_R^2 \right)-\frac{\Gamma^2}{4G\mathcal{F}} \approx
  0 \; ,\\ H_r &= \tau' P_{\tau} + R' \bar{P}_R - \Gamma {P_{\Gamma}}' \approx 0
  \; .\label{H_new_mc}\end{align}
These equations will be used as the starting point for the quantization
in Sec.~III.

We emphasize that the relative sign between the dust kinetic term
and the gravitational kinetic term can change because ${\mathcal F}>0$
($<0$) outside (inside) the horizon.
Since this change of sign is already
present in \eqref{H1},
it was not introduced by using the momentum constraint to
eliminate $P_F$. In fact, this change of sign of the gravitational part is also
present in the corresponding expression for the
Schwarzschild black hole, cf.  Eq. (117) in \cite{kuchar}.
Thus, it arises due to
the choice of the new canonical variables, which are obtained by embedding the
spatial spherically symmetric hypersurface into a LTB or Schwarzschild
spacetime. To define $F$, we had to use the canonical momentum $P_L$. Thus we
have mixed the original canonical coordinates with their momenta,
which means that the new
configuration space, spanned by $\tau$, $R$, and $F$, is different from the
original one containing the three-geometries, which is spanned by $\tau$, $R$, and
$L$. The change of sign is of fundamental interest
in the quantum theory because the original
Wheeler--DeWitt (WDW) equation has a (locally) hyperbolic kinetic term,
which is of importance for the
formulation of the proper boundary value problem \cite{OUP,Zeh}.
A change of sign has hitherto been noticed for the
WDW equation in the presence of a non-minimally coupled scalar field
\cite{Ki89}.

\subsection{Relation between dust proper time and Killing time}

In this subsection we explain under which circumstances it makes sense to
interpret $P_{\Gamma}$ as Killing time.
Equation \eqref{lost_tau} is given in the new variables by the expression
\begin{align} \label{lost_tau_prime} \tau' = 2 P_{\Gamma}' \sqrt{1+2E} +\frac{R'}{\mathcal{F}}
  \sqrt{1+2E-\mathcal{F}} \; .\end{align}
Defining $a\equiv1/\sqrt{1+2E}$ gives
\begin{align} \tau' = \frac{2P_{\Gamma}'}{a} + R' \,
  \frac{\sqrt{1-a^2\mathcal{F}}}{a\mathcal{F}} \; . \label{lost_tau_prime_2}\end{align}
If the mass density vanishes for all $r$ greater than a given $r_b$, and
if $E$ is constant for all $r$ greater than $r_b$, Eq.
\eqref{lost_tau_prime_2} can be integrated. This yields
\begin{align} a \tau &= 2 P_{\Gamma} +  \int \mathrm{d}R
  \, \frac{\sqrt{1-a^2\mathcal{F}}}{\mathcal{F}} \label{Killing_proper}\\
&= \begin{aligned}[t]2 P_{\Gamma} + F \Bigg(&
  \frac{\sqrt{1-a^2\mathcal{F}}}{1-\mathcal{F}} +
  \ln\left|\frac{1-\sqrt{1-a^2\mathcal{F}}}{1+\sqrt{1-a^2\mathcal{F}}} \right|
  \\ &\quad - \frac{1-a^2/2}{\sqrt{1-a^2}} \, \ln
  \left|\frac{\sqrt{1-a^2\mathcal{F}}-\sqrt{1-a^2}}{\sqrt{1-a^2\mathcal{F}} +
  \sqrt{1-a^2}} \right| \Bigg) \; ,\end{aligned}\end{align}
The result can also be written in the form
\begin{equation}
a\tau=2P_{\Gamma}+
F\left[\frac{\sqrt{1-a^2{\mathcal F}}}{1-{\mathcal F}}-2\tanh^{-1}\sqrt{1-a^2{\mathcal F}}+\frac{2-a^2}{\sqrt{1-a^2}}\tanh^{-1}
\frac{\sqrt{1-a^2}}{\sqrt{1-a^2{\mathcal F}}}\right]\ .
\label{int1}
\end{equation}
Here we have assumed that $0 < a \leq 1$ (corresponding to
$E\geq0$). Eq. \eqref{friedmann} guarantees that the discriminant
$1-a^2\mathcal{F}$ is non-negative. This condition may be violated
in the quantum theory, so it is of interest to give the result
for all cases. For $a>1$ but $R<Fa^2/(a^2-1)$, the result can be found
by analytic continuation to read
\begin{equation}
a\tau=2P_{\Gamma}+
F\left[\frac{\sqrt{1-a^2{\mathcal F}}}{1-{\mathcal F}}
+\ln\left|\frac{1-\sqrt{1-a^2{\mathcal F}}}{1-\sqrt{1-a^2{\mathcal F}}} \right|+
\frac{2-a^2}{\sqrt{a^2-1}}\tan^{-1}
\frac{\sqrt{a^2-1}}{\sqrt{1-a^2{\mathcal F}}}\right]\ .
\label{int2}
\end{equation}
Another analytic continuation then gives the result in the
region $R>Fa^2/(a^2-1)$,
\begin{equation}
a\tau=2P_{\Gamma}+
iF\left[\frac{\sqrt{a^2{\mathcal F}-1}}{1-{\mathcal F}}-2
\tan^{-1}\sqrt{a^2{\mathcal F}-1}+\frac{(1-a^2/2)}{\sqrt{a^2-1}}\ln
\left|\frac{\sqrt{a^2{\mathcal F}-1}-\sqrt{a^2-1}}{\sqrt{a^2{\mathcal F}-1}+\sqrt{a^2-1}}\right|\right]\ .
\label{int3}
\end{equation}
These expressions will be of relevance in the quantum theory, see Sec.~III.

We know from Birkhoff's theorem
that the spacetime around a collapsing dust cloud is given by the Schwarzschild
solution. In \cite{kuchar} it was shown for the Schwarzschild geometry that
$2P_{\Gamma}$ is equal to the Killing time $T$. Thus the equation derived
above connects the dust proper time with the Killing time at the boundary
$r_b$. For small $\Gamma$ and $E'$ the relationship may still be
used, since then we have a small amount of dust propagating in a
Schwarzschild background; only in this case the concept of Killing time makes
sense.
In the limit $a \rightarrow 1$ ($E\rightarrow 0$) we obtain
\begin{equation} \label{marginaltauT}
\tau = T + 2 \sqrt{F} \left[\sqrt{R} - \frac{\sqrt{F}}{2}
    \ln \left(\frac{\sqrt{R} + \sqrt{F}}{\sqrt{R} - \sqrt{F}} \right) \right] \; ,\end{equation}
which is identical to the relation used in \cite{VKSW03} for the
marginal case. The plus sign after the $T$ on the right-hand side
has been chosen in order to describe a collapsing dust cloud.

As has been discussed for example in \cite{MP},
Eq. \eqref{Killing_proper} gives for a Schwarzschild spacetime
a relation between Killing time and the time used by families of
freely falling observers. Each family is characterized by a fixed value
of $E$. All observers within one family start at infinity with
the speed $v_{\infty}$, where $E=v_{\infty}^2/(2(1-v_{\infty}^2))$.
In the marginal case they thus start with zero initial speed.
In this case Eq. \eqref{marginaltauT} gives the relation between
the Schwarzschild time and the Painlev\'e--Gullstrand time \cite{MP}.

\subsection{Reconstructing $E$ and $\tau_0$ from the canonical data}

The aim of this subsection is to reconstruct the local energy $E$ and the
singularity curve $\tau_0$ from the canonical data.
We have performed this reconstruction already for
the mass distribution function, see \eqref{lost_F}.
The three functions $E$, $\tau_0$, and
$F$ determine the classical LTB model completely.
There are several reasons for reconstructing these quantities. First of all,
they have a clear physical meaning, whereas this is not so evident for the
canonical coordinates. This is the reason why we have replaced the canonical
coordinates $L$ and $P_L$ with $F$ and $P_F$. One might hope that it is
possible to turn $E$ or $\tau_0$ into canonical coordinates,
although this has not been achieved yet. But even without this,
they help interpreting the canonical coordinates.

We shall first demonstrate the following connection between
the energy densities $\Gamma$ and $P_{\tau}$:
\begin{equation} \label{P_tau} P_{\tau} = \frac{\Gamma}{2\sqrt{2E+1}} \;
  \end{equation}
for an arbitrary foliation. Consider the momentum constraint
\begin{align} 0 &= \tau' + \frac{R' \bar{P}_R}{P_{\tau}} +
  \frac{P_F \Gamma}{P_{\tau}} \\ &= \tau' - \frac{R'}{\mathcal{F}}
  \sqrt{\frac{\Gamma^2}{4P_{\tau}^2}-\mathcal{F}} + \frac{P_F
    \Gamma}{P_{\tau}} \label{supi} \; ,\end{align}
where, in the second step,
we have used the Hamiltonian constraint \eqref{H_new} in the form
\begin{equation} \label{P_R_2} \bar{P}_R
  = - \frac{P_{\tau}}{\mathcal{F}} \sqrt{\frac{\Gamma^2}{4
      P_{\tau}^2}-\mathcal{F}} \; . \end{equation}
Comparing \eqref{supi} with \eqref{lost_tau_prime} we can read off Eq.
\eqref{P_tau}, which is also a constraint.
  Equation \eqref{P_tau} then gives us a simple expression
for $E$ in terms of the canonical variables,
\begin{equation} 1+2E =  \frac{\Gamma^2}{4 P_{\tau}^2}
  \stackrel{\makebox[0cm]{\scriptsize \eqref{H_new}}}{=} -
  \frac{\Gamma^2}{4\mathcal{F}\bar{P}_R^2 - \Gamma^2/(\mathcal{F})} \; .
 \label{E_p_r}\end{equation}
Note that, using the solution \eqref{ltb-sol} of Einstein's equation, we have an expression for
$\tau_0-\tau$, that is, for the remaining proper dust time until the dust shell reaches the singularity,
\begin{align} \label{tau_zero}\tau_0-\tau =
  \frac{R^{3/2}Q\left(-\frac{2ER}{F}\right)}{\sqrt{F}} \; .\end{align}
Since we know now how to express $E$ by canonical coordinates, this equation
gives us an expression for the bang time $\tau_0$ in
terms of canonical coordinates.

Now one can express $\bar{P}_R$ in terms of $F$ and $E$. Inserting the
expression for $P_{\tau}$ into \eqref{P_R_2}, we have
\begin{align} \bar{P}_R &= \frac{1}{\mathcal{F}} \, \frac{\Gamma}{2\sqrt{2E+1}} \,
  (\pm1)\sqrt{2E+1-\mathcal{F}} \\ &=\frac{1}{\mathcal{F}} \,
  \frac{\Gamma}{2\sqrt{2E+1}} \, \partial_{\tau}R \; .\end{align}

\subsection{Hamiltonian equations of motion}

Here we shall give the Hamilton equations of motion and derive
Einstein's equation \eqref{ltb-eg} from them.
The Hamiltonian equations are generally given by
\begin{align} \dot{X} &= \{ X, \mathcal{H}[N] + \mathcal{H}_r[N^r] \} \; ,\\ \dot{P}_X &= \{P_X,\mathcal{H}[N]
  + \mathcal{H}_r[N^r] \} \; ,\end{align}
where we have introduced the smeared constraints
\begin{align} \label{super-ham-pb} \mathcal{H}[N] = \int_0^{\infty} \,
\mathrm{d}r \, N(r) \, H(r) \, , \quad \mathcal{H}_r[N^r] = \int_0^{\infty} \, \mathrm{d}r \,
N^r(r) \, H_r(r) \; .\end{align}
Starting from the action \eqref{action_2},
the Hamiltonian equations of motion are\footnote{Note
  that $\delta F(r) / \delta \Gamma(\bar{r}) = \theta(r-\bar{r})$.}
\begin{align}
  \dot{\tau} &=  2 N P_{\tau} + N^r \tau' \label{dottau} \; ,\\
  \dot{P}_{\tau} &= (N_r P_{\tau})' \; ,\\ \dot{R} &= 2 N \mathcal{F} \bar{P}_R + N_r R'
  \; ,\\ \dot{\bar{P}}_R &=  - N \left( \frac{F \bar{P}_R^2}{R^2} + \frac{\Gamma^2 F }{4
  \mathcal{F}^2 R^2} \right) + (N^r \bar{P}_R)' \; ,\\ \dot{\Gamma} &= (N^r
  \Gamma)' \label{dot_gamma}\; ,\\ \dot{P}_{\Gamma} &= N \frac{\Gamma}{2 \mathcal{F}} + N^r
  P_{\Gamma}' + \int_r^{\infty} \mathrm{d}\tilde{r} \, N(\tilde{r})
  \left(\frac{\bar{P}_R^2(\tilde{r})}{R(\tilde{r})} + \frac{\Gamma^2(\tilde{r})}{4
  \mathcal{F}^2(\tilde{r}) R(\tilde{r})} \right) \; . \end{align}
 Consider now
\begin{align}\partial_{\tau} R = \frac{\dot{R}}{\dot{\tau}}
  \stackrel{N^r=0}{=} \frac{ N 2 \mathcal{F} \bar{P}_R}{N2P_{\tau}}
  \stackrel{\eqref{P_tau}}{=} \frac{\mathcal{F}\bar{P}_R2\sqrt{2E+1}}{\Gamma}
  \; ,  \end{align}
which we can solve for $\bar{P}_R$. Inserting this expression into the
Hamiltonian constraint \eqref{H_new} and again using \eqref{P_tau} gives
\begin{equation}0 = H = \frac{\Gamma^2}{4(2E+1)} + \frac{\Gamma^2
    (\partial_{\tau}R)^2}{\mathcal{F} 4 (2E+1)} - \frac{\Gamma^2}{4
    \mathcal{F}} \; .\end{equation}
Solving for $(\partial_{\tau}R)^2$ leads to Einstein's equation
\eqref{ltb-eg}.
Note that we did not have to specify the lapse function.

One can show, moreover, that $F$, $E$, and $\tau_0$
are constants of motion, that is,
they have vanishing Poisson brackets with the Hamiltonian constraint.
In their given local form they do, of course, not commute with the
momentum constraint, since the latter generates their transformation
with respect to a relabelling $r\to f(r)$.
One would, however, expect that a suitable non-local form 
commutes with $H_r$ and thus turns them into real `observables'
(similar to what one would expect to happen with
 the geometric operators in loop quantum gravity \cite{OUP}).
This would coincide with the interpretation of them being the
physically relevant variables energy and bang time.

We also note that because of \eqref{dottau},
%
%
$N$ has no longer (for vanishing shift) the interpretation of being the ratio of proper time to
ADM time, cf. \cite{OUP}.
 The reason is that we have squared the original version of the
Hamiltonian constraint, see App.~A.
If we define a new version of the constraint by taking a square root,
\begin{equation} \label{H_uparrow} H_{\uparrow} =
  P_{\tau} - \sqrt{-\mathcal{F} \bar{P}_R^2 + \frac{\Gamma^2}{4G^2\mathcal{F}}}
\approx 0 \; ,\end{equation}
we find that
\begin{align} \{\tau, \mathcal{H}_{\uparrow}[N^{\uparrow}]\} = N^{\uparrow}
   \,   \label{pb1} \end{align}
and recover for the lapse function $N^{\uparrow}$ the old interpretation.

We finally remark that the algebra of the constraints cannot be
of the general form (given for example in \cite{OUP}), because we have used
the momentum constraint to eliminate $P_F$ in the Hamiltonian constraint.
In fact, a short calculation gives
\begin{align} \left\{\mathcal{H}[N], \, \mathcal{H}[M]\right\} &= 0 \; , \label{ham_poiss}\\
  \left\{\mathcal{H}_r[N^r], \, \mathcal{H}[N]\right\} &= \mathcal{H}[N_{,r}
  N^r - N N_{,r}^r] \; , \label{ham_1}\\ \left\{\mathcal{H}_r[N^r], \,
    \mathcal{H}_r[M^r]\right\} &= \mathcal{H}_r\left[[N^r,M^r]\right] \; .\label{ham_2}
\end{align}
We note that the Poisson bracket of the Hamiltonian with itself
vanishes, Eq. \eqref{ham_poiss},
in contrast to the general case where it closes on the
momentum constraint. The other brackets coincide with the general case.
The transformations generated by the Hamiltonian
constraint can thus no longer be interpreted as hypersurface deformations.
They are in general not orthogonal to the hypersurfaces, but act
along the flow lines of dust.


\section{Diffeomorphism invariant quantum states}

\subsection{Quantum constraints}

We shall now apply the quantization procedure proposed by Dirac and turn
the classical constraints into quantum operators, cf. \cite{OUP}.
Starting point are thus the expressions
\eqref{H_new} and \eqref{H_new_mc}.

The translation of Poisson brackets into commutators is achieved in
the Schr\"odinger representation by substituting
\begin{align} P_{\tau}(r)
\rightarrow \frac{\hbar}{i} \,\frac{\delta}{\delta \tau(r)} \, , \quad
\bar{P}_{R}(r) \rightarrow \frac{\hbar}{i} \,\frac{\delta}{\delta R(r)} \, ,
\quad P_{\Gamma}(r) \rightarrow \frac{\hbar}{i} \,\frac{\delta}{\delta
  \Gamma(r)}  \label{prescription}\end{align}
and acting with them on wave functionals.
The Hamiltonian constraint \eqref{H_new} then leads to the
WDW equation,
\begin{align} \begin{split} \label{WDW_eq}\Bigg[- G \hbar^2 \Bigg(
    \frac{\delta^2}{\delta \tau(r)^2} & + \mathcal{F} \,
    \frac{\delta^2}{\delta R(r)^2} + A(R,F) \, \delta(0) \,
    \frac{\delta}{\delta R(r)} \\ & + B(R,F) \, \delta(0)^2 \Bigg) -
    \frac{\Gamma^2}{4G\mathcal{F}} \Bigg]
    \Psi\left[\tau(r'),R(r'),\Gamma(r')\right] = 0 \; ,\end{split} \end{align}
where $A$ and $B$ are smooth functions of $R$ and $F$ that encapsulate the
factor ordering ambiguities. We have introduced
divergent quantities such as $\delta(0)$ in order to indicate that
the factor ordering problem is unsolved and can be dealt with only
after some suitable regularization has been performed, cf. \cite{TW87}.
That is, one would like to choose the terms proportional to
$\delta(0)$ in such a way that the constraint algebra closes, which is usually
called `Dirac consistency'.

Quantizing the momentum constraint \eqref{H_new_mc} by using \eqref{prescription} gives
\begin{align} \left[\tau' \frac{\delta}{\delta \tau(r)}  + R' \frac{\delta}{\delta
      R(r)} - \Gamma \left(\frac{\delta}{\delta \Gamma(r)}\right)' \right]
  \Psi\left[\tau(r'), R(r'), \Gamma(r') \right] = 0 \; . \end{align}
Up to now, the quantum constraint equations have been formulated
only in a formal way. The next subsection is devoted to the application
of a lattice regularization.

\subsection{Lattice regularization}

We follow here the suggestion made in \cite{VWS04} and consider
a one-dimensional lattice given by a discrete set of points $r_i$
separated by a distance $\sigma$. In order that the momentum constraint
is fulfilled in the continuum limit, it is important to start
with a corresponding ansatz for the wave functional before putting
it on the lattice. We therefore make the ansatz
\begin{align} \Psi\left[\tau(r), \, R(r), \, \Gamma(r)\right] =
U \left( \int \mathrm{d}r \, \Gamma(r) \mathcal{W}
(\tau(r), \, R(r),\, \Gamma(r))\right) \; ,\end{align}
where $U: \mathbb{R} \rightarrow\mathbb{C}\,$ is at this stage some
arbitrary (differentiable) function.
Using $\Gamma$ in the exponent instead of $R'$ or $\tau'$ is
suggested by the form of the WDW equation (absence of derivatives with respect
to $\Gamma$) and the fact
that $F'=\Gamma$ is related to the energy density.
The ansatz has to be compatible
with the lattice, which means that it has to factorize into different functions
for each lattice point. So we have to make the choice $U=\exp$, which gives
\begin{align} \lefteqn{\Psi\left[\tau(r), \, R(r), \,
    \Gamma(r)\right]} \\ \label{ansatz_cont_gamma} &= \exp \left( \int \mathrm{d}r \, \Gamma(r)
  \mathcal{W}(\tau(r), \, R(r),\, F(r)) \right) \\ \label{ansatz_limit_gamma}&=
\exp \left( \lim_{\sigma \rightarrow 0} \sum_i \sigma
  \Gamma_i \mathcal{W}_i\left(\tau(r_i),\, R(r_i), \,
    F(r_i)\right) \right) \\ \label{ansatz_latt_gamma} &= \lim_{\sigma \rightarrow
  0} \, \prod_i \,\exp \left( \sigma \, \Gamma_i \,
  \mathcal{W}_i\left(\tau(r_i),\, R(r_i), \, F(r_i)\right) \right) \\ &=
\label{factorize_gamma} \lim_{\sigma \rightarrow 0} \,\prod_i \,
\Psi_i\left(\tau(r_i), \, R(r_i), \, \Gamma(r_i), \,  F(r_i)\right) \; ,\end{align}
where
\begin{equation} F(r_i) = \sum_{j=0}^i \, \sigma \,\Gamma_j \; .\end{equation}
As in \cite{VWS04} we implement the formal expression $\delta(0)$
onto the lattice as follows,
\begin{align} \delta(0) \rightarrow \lim_{\sigma \rightarrow 0}
  \frac{1}{\sigma} \; . \end{align}
The lattice version of the WDW equation \eqref{WDW_eq} then reads,
\begin{equation}
 \label{WDW_lattice_gamma}\Bigg[
      G\hbar^2\left(\frac{\partial^2}{\partial \tau_j^2}  + \mathcal{F}_j \,
    \frac{\partial^2}{\partial R_j^2} + A(R_j, F_j)  \,
    \frac{\partial}{\partial R_j}\right) + B(R,F) +
    \frac{\sigma^2 \, \Gamma}{4G \, \mathcal{F}_j}\Bigg] \Psi_j =
    0 \; .\end{equation}
We  now insert the ansatz \eqref{ansatz_latt_gamma} and make for
convenience the redefinition $\mathcal{W}=iW/2$. This leads to
\begin{multline} \frac{\sigma^2 \Gamma_i^2}{4}
  \left[G\hbar^2\left(\frac{\partial W(\tau,\, R,\, F)}{\partial
        \tau}\right)^2+G\hbar^2\mathcal{F} \, \left( \frac{\partial W(\tau, \,R,\,
        F)}{\partial R} \right)^2 - \frac{1}{G\mathcal{F}} \right] \\ + \;
        \frac{\sigma \Gamma_i}{2}
  \left[ G\hbar^2\left(\frac{\partial^2}{\partial \tau^2} + \mathcal{F} \,
      \frac{\partial}{\partial R^2} + A(R,F) \frac{\partial}{\partial R}
    \right) W(\tau, \, R, \, F) \right] \\ + B(R,F) = 0 \; . \end{multline}
In order for this to be fulfilled independent of the choice of $\sigma$
(and thus also in the limit $\sigma\to 0$) one is led to the following
{\em three} equations,
\begin{align} \left(G\hbar\frac{\partial W(\tau, \, R, \, \Gamma)}{\partial \tau}
  \right)^2 + \mathcal{F} \, \left(G\hbar \frac{\partial W(\tau, \, R, \,
      \Gamma)}{\partial R} \right)^2 - \frac{1}{\mathcal{F}} & = 0 \label{W1_old} \; , \\
  \left(\frac{\partial^2}{\partial \tau^2} + \mathcal{F} \,
    \frac{\partial^2}{\partial R^2} + A(R,\Gamma) \frac{\partial}{\partial R}
  \right) W(\tau, \, R, \, \Gamma) & = 0 \; ,
\label{W2_old} \\ \intertext{and} B(R,\Gamma) = 0 \; . \label{B_old}\end{align}
The first equation, \eqref{W1_old}, the Hamilton-Jacobi equation,
is the same as
in \cite{VWS01}, Eq.  (5.11) and in \cite{VKSW03}, Eq. (12).
The second equation
presents an additional restriction on solutions of \eqref{W1_old}.
The last equation
\eqref{B_old} tells us that working on the lattice is only possible if the
factor ordering does not contribute to the potential term. If we find
solutions to all three equations, we can do all other calculations on the lattice,
since these solutions have a well defined continuum limit and satisfy the
momentum constraint.

At this point we have to make a few comments on the regularization procedure
we are using. It was already noted that the lattice regularization does not solve the factor ordering
problem. The lattice regularization just represents an ad hoc regularization in which the divergent
terms have to cancel each other. Put differently, it is equivalent to a DeWitt
type of regularization (which means setting $\delta(0)=0$) with an additional
constraint on the solutions. It has to be noted that
the Hamilton--Jacobi equation, Eq. \eqref{W1_old},
can equivalently be obtained from the highest order of a WKB approximation
\cite{VKSW03}.

We have already emphasized that the signature in the
kinetic part of the
Hamiltonian constraint \eqref{H_new} can change from elliptic
(outside the horizon) to hyperbolic (inside the horizon).
This thus occurs for the kinetic term of the
WDW equation \eqref{WDW_eq}, too. As already noted
above, a similar phenomenon can be found in the
context of quantum cosmology \cite{Ki89}.
As discussed in \cite{BK97}, we can say that
the part inside the horizon is always classically allowed,
whereas this is not necessarily the case for the outside part.
The usual initial value problem appropriate for hyperbolic equations
can thus only be applied for the region corresponding to the
black hole interior.

Since we have found from our special ansatz the two equations
\eqref{W1_old} and \eqref{W2_old}, we cannot expect for them
a well posed initial value problem to hold because the system is in
general overdetermined. The goal pursued in this paper is not
the discussion of boundary value problems but to find a class of
exact solutions to all quantum constraints and to draw from them
physical conclusions.

It is instructive to check that one can get Einstein's equation \eqref{ltb-eg}
from the Hamilton--Jacobi equation \eqref{W1_old}.
A class of solutions is given by the complete integral
\begin{align}\label{Ham_Jac_sols}
G\hbar\, W(\tau, \, R, \, a, \, b) = b + a\tau \pm \int \mathrm{d}R
  \frac{\sqrt{1-a^2 \mathcal{F}}}{\mathcal{F}} \; ,\end{align}
which is identical to the expression occurring in
\eqref{Killing_proper}, cf. also Eq. (3.3) in
\cite{MP}.
Since we made an ansatz of the form
$\Psi \equiv \exp(i S/G\hbar) \equiv \exp( i \int (F'/2) \,
W)$, the relation
\begin{align} P_R \; &= \; \frac{F'}{2} \, \frac{\partial}{\partial R} W  \\
  &= \; \frac{F'}{2} \; \frac{\sqrt{1-a^2 \mathcal{F}}}{\mathcal{F}} \label{consis_1}\end{align}
should hold. Using the Hamiltonian equation of motion for zero shift we have
\begin{equation} \dot{R} = 2 N \mathcal{F} \, P_R \; .\end{equation}
Inserting this in \eqref{consis_1} gives
\begin{align} \label{consis_2}\left( \frac{\dot{R}}{N F'} \right)^2 = 1 - a^2 + a^2
  \frac{F}{R} \; . \end{align}
This equation should be equivalent to Einstein's equation given by
\begin{align} \left(\frac{\mathrm{d}R}{\mathrm{d}\tau}\right)^2 =
  \frac{F}{R} + 2E \; .\end{align}
For $a=0$ the right-hand side of \eqref{consis_2} is equal to $1$.
This is a limiting case. As discussed in \cite{MP}, it corresponds to
observers that start at ${\mathcal J}^-$ using the Eddington--Finkelstein
coordinate $v$ as their time coordinate. For an arbitrary
$a \neq 0$ we can write
\begin{align} \left( \frac{\dot{R}}{N F'a} \right)^2 = \frac{1}{a^2} - 1 +
  \frac{F}{R} \end{align}
and can thus identify $2E= 1/a^2 - 1$. Hence, $a$ has the same interpretation as
in \eqref{Killing_proper}. Letting $a$ go to zero implies that $E$ diverges.

\subsection{Solutions}

We have obtained above the two equations \eqref{W1_old}
and \eqref{W2_old}, which have to be satisfied in order to get a diffeomorphism
invariant solution to the WDW equation.  (We set now $G=1=\hbar$.)
Looking for particular solutions of
the separating
form $W=\alpha(\tau)+\beta(R)$, we recognize immediately that this system
of equations is consistent {\em only for special factor orderings}.
The particular
factor ordering $A(R,F)=F/(2R^2)$ used in \cite{VKSW03} and
\cite{VWS04} is not among them. (For the WKB solutions used in
\cite{VKSW03} this is irrelevant.)

Tackling the problem from the opposite point of view,
one can ask for which factor orderings
we {\em do} get a separating solution. We find
\begin{align}  A(R,F) &= \frac{F}{2R^2} \; \left(1 +
\frac{1}{1-a^2 \mathcal{F}} \right)\; . \label{uni_factor}\end{align}
This leads to
\begin{align} \label{W-gen} W(\tau,\, \Gamma,\, R, a) &= \mathrm{const.} \pm  a
  \tau \pm  \,\int dR \; \frac{\sqrt{1-a^2\mathcal{F}}}{\mathcal{F}} \;
  .\end{align}
These are identical to the solutions that were obtained in
the solution of the Hamilton--Jacobi equation,
cf. \eqref{Ham_Jac_sols} and \eqref{int1}. Thus we can again
identify $2E=1/a^2-1$. Since classically $E
\geq -1/2$, it follows that $a$ should be real.
Using \eqref{P_tau}, we can show
that this is consistent:
\begin{align}\label{P_tau_con} \hat{P}_{\tau} \Psi_a = \pm \frac{a \,
    \Gamma}{2} \Psi_a = \pm \frac{\Gamma}{2\sqrt{2E+1}} \Psi_a \;
  ,\end{align}
where we have defined
\begin{equation} \Psi_a[\tau, R, \Gamma] = \exp\left( \frac{i}{2}\int_0^{\infty}
    \mathrm{d}r \, \Gamma W(\tau, R, F, a) \right)\; . \label{all_solutions}\end{equation}
The integral appearing in \eqref{W-gen} has been evaluated above for the
various cases, see \eqref{int1}, \eqref{int2}, \eqref{int3}.
 We recognize in particular
from \eqref{int3} that the wave function becomes a real exponential in the
region $R<Fa^2/(a^2-1)$. As can be seen from \eqref{friedmann},
this is the region that is classically forbidden; the real, non-oscillatory,
behaviour of the wave function is thus no surprise.

Surprisingly, Eq. \eqref{W-gen} gives in fact already the {\em complete} class
of solutions. This is shown in Appendix~B. There thus exist no
non-separating solutions, given the ansatz made where the full wave functional
factorizes into functions on the respective lattice points.
But we emphasize that we have succeeded in finding exact solutions
to all quantum constraints. Other solutions to the full WDW equation
and momentum constraints would necessarily couple the infinitely many shells
comprising the dust cloud; to find them would demand a regularization
scheme that is much beyond the scope of this paper.
In a sense, the factor ordering chosen here leads to quantums states
for which the WKB form is ``exact'', cf. the analogous situation with
the models discussed in \cite{BK97} and \cite{LGK}.

For definiteness we will consider in the following only the
positive sign in front of $a\tau$ in \eqref{W-gen}.
For $a=0$, which corresponds to $E=\infty$, the solutions are
particularly simple. They read
\begin{align}
\label{W-0-in}W_{\mathrm{in}}^{\pm}(\tau,\, \Gamma, \, R, a=0) &= \mathrm{const.} \pm  \left( R + F
  \ln(F-R) \right) \\ \label{W-0-out}W_{\mathrm{out}}^{\pm}(\tau,\, \Gamma, \, R, a=0)
&= \mathrm{const.} \pm  \left( R + F \ln(R-F) \right) \; ,\end{align}
where `in' refers to the region inside the horizon and `out' to the region
outside the horizon.
For $a=1$, which corresponds to the marginal model $E=0$, we obtain the solutions
\begin{align} \label{sol_E_zero} W_{\mathrm{in}}^{\pm}(\tau,\Gamma,R,a=1) &= \mathrm{const.} +
   \tau \pm 2 \sqrt{F} \left[\sqrt{R} - \sqrt{F}
    \tanh^{-1}\left(\sqrt{\frac{R}{F}}\right) \right] \\ \label{sol_E_zero_2}
  W_{\mathrm{out}}^{\pm}(\tau,\Gamma,R,a=1) &= \mathrm{const.} +  \tau \pm 2 \sqrt{F}
  \left[\sqrt{R} - \sqrt{F} \tanh^{-1}\left(\sqrt{\frac{F}{R}}\right)
  \right].\end{align}
These solutions were already found in \cite{VWS01} and discussed further in
\cite{VKSW03}.

The full solutions for the marginal model read
\begin{equation} \Psi_{\mathrm{in/out}}^{\pm}[\tau, R, \Gamma] =
\Psi_{\Gamma\,\mathrm{in/out}}[\Gamma] \,
  \exp\left( \frac{i}{2} \,\int_0^{\infty} \mathrm{d}r \, \Gamma(r)
  W_{\mathrm{in/out}}^{\pm}(\tau,R,\Gamma,a=1) \right)\; . \end{equation}
The solutions $\Psi_{\mathrm{in}}^{\pm}$ and $\Psi_{\mathrm{out}}^{\pm}$ have
to be matched at the horizon, that is, both the states and their
$R$-derivatives should agree there. As can be seen from \eqref{sol_E_zero}
and \eqref{sol_E_zero_2}, however, the phases of the states diverge there.
One thus has to perform an analytic continuation: We write
$R-F=\epsilon\exp(i\varphi)$, $\epsilon>0$, and compare the states
at $\varphi=\pi/2$, that is, at $R=R_h\equiv F+i\epsilon$.
One finds from this comparison that the states are related as
\begin{equation} \Psi_{\mathrm{in}}^{\pm}[\tau, R_h, \Gamma]
  \; = \frac{\Psi_{\Gamma\, \mathrm{in}}[\Gamma]}
{\Psi_{\Gamma\, \mathrm{out}}[\Gamma]}
\; \exp\left(\mp \frac{\pi}{2} \int_0^{\infty} \mathrm{d}r \, \Gamma(r) \,
    F(r) \right) \; \Psi_{\mathrm{out}}^{\pm}[\tau, R_h, \Gamma] \; .\end{equation}
The states $\Psi_{\mathrm{in}}^{\pm}$ and $\Psi_{\mathrm{out}}^{\pm}$
can be set equal at the horizon if we exploit the freedom to choose
$\Psi_\Gamma[\Gamma]$. One thus gets a relation between
$\Psi_{\Gamma\,\mathrm{in}}^{\pm}[\Gamma]$ and
$\Psi_{\Gamma \,\mathrm{out}}^{\pm}[\Gamma]$,
\begin{align} \Psi_{\Gamma\,\mathrm{in}}^{\pm}[\Gamma] \; &= \; \exp\left(\mp \frac{\pi}{2}
    \int_0^{\infty} \mathrm{d}r \, \Gamma(r) \, F(r) \right) \; \Psi_{\Gamma \,
    \mathrm{out}}^{\pm}[\Gamma] \\ &= \; \exp\left(\mp \frac{\pi}{4}
    (F^2(\infty)-F^2(0))\right) \; \Psi_{\Gamma \, \mathrm{out}}^{\pm}[\Gamma] \\ &= \;
  \exp\left(\mp \pi \, M^2 \right) \; \Psi_{\Gamma \,
    \mathrm{out}}^{\pm}[\Gamma]\;.\label{entropy} \end{align}
On the lattice this reads
\begin{equation} \Psi_{\Gamma \,\mathrm{in}}^{\pm}(\Gamma_i, \, F_i) \; = \;
  \exp\left(\mp \pi \omega_i \, F_i \right) \; \Psi_{\Gamma \,
    \mathrm{out}}^{\pm}(\Gamma_i,\, F_i) \; . \end{equation}
One can easily check that with this choice the derivatives
of the states at the horizon coincide as well. We also remark that
the alternative choice $R=F-i\epsilon$ would lead to a switch of sign
in the exponent of \eqref{entropy}.

We mention that the factor acquired when crossing the horizon might be
connected to the entropy of the black hole \cite{BK97}.
In fact, we recognize that the absolute value of the exponent
is one quarter of the Bekenstein--Hawking entropy.

The solutions \eqref{W-gen} are all solutions
that can occur on the lattice for the factor ordering
\eqref{uni_factor}. Here we want to extend these solutions to the continuum.
The solutions on the lattice contain two free parameters, $a_i$ and $b_i\,$,
\begin{equation}\Psi_i = e^{ib_i \sigma \Gamma_i/2} \, e^{\frac{i}{2} \sigma
    \Gamma (a_i\tau_i + \int \mathrm{d}R
    \frac{\sqrt{1-a_i\mathcal{F}_i}}{\mathcal{F}_i})} \; .\end{equation}
(We have for simplicity considered only one sign in the exponent.)
We know that $a_i$ is connected with the local energy $E$ via $2E_i+1 =
1/a_i^2$. In general we have $E=E(r)$, and thus it would be natural to demand
that $a=a(r)$. But with this explict dependence on $r$
the momentum constraint would not be satisfied. The
only possible way out of this is to use an implicit dependence
$a(F(r))$ in order to get solutions fulfilling the momentum constraint. The
expression
\begin{align}\Psi[\tau, R, \Gamma] =  &e^{(i/2)\int \mathrm{d}r \, b(F(r)) \Gamma } \,
  \notag\\ &\times \exp\left\{\frac{i}{2} \int \mathrm{d}r \,
    \Gamma \;\left[ a(F(r))\tau + \int \mathrm{d}R
    \frac{\sqrt{1-a(F(r))\mathcal{F}}}{\mathcal{F}}\right] \right\} \; \label{meg_sol}\end{align}
is still a solution, since the Hamiltonian constraint does not contain a
derivative with respect to $\Gamma$ or $F$. Hence we arrive at a family of
solutions containing two arbitrary functions $a(x)$ and $b(x)$. It is clear
that $a(x)$ is connected with $E$ via $2E(r)+1 = 1/a^2(F(r))$.
One might in principle wish to construct wave packets by superposing
wave functions with {\em different} $a$, that is, with different energies.
However, since the factor ordering depends  on $E$, this does not seem
feasible.


\section{Discussion}

Let us first summarize the main results of our paper.
As for the classical part, we have extended the canonical formalism to
the non-marginal case. We have shown in particular that \eqref{H_new}
holds for all values of $E$. We have succeeded in expressing
the observables $F$, $E$, and $\tau_0$, which determine the
classical model completely, in terms of the canonical variables.
We have presented a new and improved method to cope with the
boundary action in the canonical formalism. This renders the formalism
more transparent. We have presented the general relation between
dust proper time and Killing time and noted the similarity of dust time with
the generalized Painlev\'e--Gullstrand time discussed in \cite{MP}.

As for the quantum part, we have presented a lattice regularization
that correctly implements the momentum constraint in the continuum limit.
For a particular factor ordering (that, however, depends on $E$)
we have succeeded
to give exact solutions to all constraints. We have shown that, given
the general ansatz for the wave functional on the lattice, these are
the only solutions. This means that these are the only solutions
for which the states describing the dust cloud factorize into infinitely
many states corresponding to the various shells forming the cloud.
We have also extended these solutions into the continuum.

The discussion of the quantum LTB model is far from being exhausted.
We thus want to
conclude with an outlook on future work. First, we have mentioned
in Sec.~III.C a possible connection between the exponential factor in
\eqref{entropy} and the Bekenstein--Hawking entropy. Perhaps it will be
possible to recover this entropy as an entanglement entropy from
these quantum states. This could provide a first step towards
its general understanding. Second, it would be of great importance
to study the possible singularity avoidance of the quantum LTB model.
This would include also an understanding of the role of the
naked singularities in the classical model. We recall that singularity
avoidance was a main feature of the quantization of dust shells
\cite{hajicek}. Singularity avoidance was also shown in various models
of loop quantum gravity \cite{loopqg}. A third issue would thus
be to develop the loop quantization of the LTB model and compare its
results with the results obtained from the WDW equation.

Extending the discussion of Hawking radiation in \cite{VKSW03,VWS04},
we plan to derive the corresponding Bogoliubov coefficients from
the exact quantum states found in this paper. This should yield
in the appropriate limits the thermal Hawking spectrum plus
quantum gravitational correction terms. This is currently under investigation.

\section*{Acknowledgements}

\noindent We thank T. P. Singh, Carsten Weber, L. C. R. Wijewardhana,
 and Louis Witten
for discussions. C. K. is grateful to the Max--Planck-Institut f\"ur
Gravitations\-physik in Golm, Germany, for its kind hospitality while part
of this work was done.


\begin{appendix}

\section{Simplification of the Hamiltonian constraint}

Here we show how it is possible to eliminate $P_F$ from \eqref{H1} by using
\eqref{H2} and to obtain the new, relatively simple, version \eqref{H_new} of
the Hamiltonian constraint.

The momentum
constraint \eqref{H2} can be solved for $\bar{P}_R$, which gives
\begin{equation} \bar{P}_R = - \frac{\tau' P_{\tau} - \Gamma P_{\Gamma}'}{R'} \; .\end{equation}
Inserting this expression into the Hamiltonian constraint \eqref{H1}
and noting that $L^2 = {R'}^2/\mathcal{F} - 4 \mathcal{F} P_{\Gamma}'^2$ gives
\begin{align} 2P_{\tau} \sqrt{L^2+{\tau'}^2} &= \left[\frac{\Gamma
R'}{\mathcal{F}} + \frac{4 \mathcal{F} P_{\Gamma}'}{R'} \left( \tau'
P_{\tau} - \Gamma P_{\Gamma}'\right) \right] \\ &= \frac{\Gamma L^2 + 4
\mathcal{F} \tau' P_{\Gamma}' P_{\tau}}{R'} \; .\end{align}
Squaring the resulting equation yields
\begin{align} 4R'^2 P_{\tau}^2 L^2 + 4 R'^2 P_{\tau}^2 \tau'^2 &= \Gamma^2 L^4 + 8 \mathcal{F}
 L^2 \Gamma \tau' P_{\Gamma}' P_{\tau} + \underbrace{16 \mathcal{F}^2
 \tau'^2 P_{\Gamma}'^2 P_{\tau}^2}_{ 4 \tau'^2 P_{\tau}^2
 \left(R'-\mathcal{F}L^2\right)} \; .\end{align}
The second term on the left cancels with the corresponding term on
 the right. After dividing by $L^2$ one has
\begin{align} 4R'^2 P_{\tau}^2 + 4 \mathcal{F} P_{\tau}^2 \tau'^2 = \Gamma^2
(\underbrace{R'^2/\mathcal{F} -4 \mathcal{F} P_{\Gamma}'^2}_{L^2}) + 8
\mathcal{F} \Gamma \tau' P_{\Gamma}' P_{\tau} \; .\end{align}
If we combine the second term on the left-hand side with the second and
third term on the right-hand side, it is possible to use the momentum
constraint to eliminate $P_{\Gamma}'$,
\begin{align}4 R'^2 P_{\tau}^2 = \frac{\Gamma^2 R'^2 }{\mathcal{F}}- 4\mathcal{F}
\underbrace{(\tau' P_{\tau} - \Gamma
P_{\Gamma}')^2}_{R'^2\bar{P}_R^2} \; .\end{align}
This then yields the desired expression for \eqref{H_new}:
\begin{equation}P_{\tau}^2+ \mathcal{F} \bar{P}_R^2-\frac{\Gamma^2}{4\mathcal{F}} = 0 \; . \end{equation}

\section{Uniqueness of quantum solutions}

Here we shall demonstrate that the separating solutions found
from \eqref{W1_old} and \eqref{W2_old} are unique.

Equation \eqref{W1_old} may be solved by the ansatz
\begin{align}\partial_{\tau}W = \frac{\cos \eta}{\sqrt{\mathcal{F}}} \, ,\quad
  \quad \partial_R W= \frac{\sin \eta}{\mathcal{F}} \;. \label{u_ansatz}\end{align}
The function $\eta$ has to fulfill the integrability condition
\begin{align} -\frac{\sin \eta}{\sqrt{\mathcal{F}}} \partial_R\eta -
  \frac{\cos \eta}{2 \mathcal{F}^{3/2}} \partial_R\mathcal{F} &= \frac{\cos
  \eta}{\mathcal{F}} \partial_{\tau} \eta \; ,\end{align} which leads to
\begin{align}\partial_{\tau} \eta = - \sqrt{\mathcal{F}} \tan \eta \partial_R
  \eta - \partial_R \sqrt{\mathcal{F}} \; .\label{u_int}\end{align}
Inserting ansatz \eqref{u_ansatz} into \eqref{W2_old} gives another equation
for $\eta$,
\begin{align}- \frac{1}{\sqrt{\mathcal{F}}} \partial_{\tau}\eta + \cot \eta
  \partial_R \eta - \partial_R \ln \mathcal{F} + \frac{A}{\mathcal{F}} = 0 \;
  .  \end{align}
So the ansatz $A=\mathcal{F} \partial_R \ln(\mu\mathcal{F})$ yields
\begin{align} \partial_{\tau} \eta = \sqrt{\mathcal{F}} \cot \eta \partial_R
  \eta + \sqrt{\mathcal{F}} \partial_R \ln \mu \; . \label{u_w2}\end{align}
The integrability conditions \eqref{u_int} and \eqref{u_w2} have to be
consistent. This gives
\begin{align} - \sqrt{\mathcal{F}} \tan \eta \partial_R
  \eta - \partial_R \sqrt{\mathcal{F}} =\sqrt{\mathcal{F}} \cot \eta \partial_R
  \eta + \sqrt{\mathcal{F}} \partial_R \ln \mu \: .\end{align}
By an elementary manipulation we obtain
\begin{align} \partial_R \ln\left(\sqrt{\mathcal{F}} \mu \tan \eta \right) = 0
  \; ,\end{align}
and thus
\begin{align} \tan \eta = \frac{\alpha(\tau)}{\mu \sqrt{\mathcal{F}}}, \quad
  \sin \eta = \frac{\alpha(\tau)}{\sqrt{\alpha^2 + \mu^2 \mathcal{F}}}, \quad
  \cos \eta = \frac{\mu \sqrt{\mathcal{F}}}{\sqrt{\alpha^2 + \mu^2
  \mathcal{F}}} \; .\end{align}
These equations can be used to give expressions for
 $\partial_R \eta$ and $\partial_{\tau}\eta$,
\begin{align} \partial_R \eta &= \frac{\alpha \mu^2 \mathcal{F}}{\alpha^2+\mu^2
    \mathcal{F}} \partial_R \left( \frac{1}{\mu \mathcal{F}} \right)  \label{p_r_eta}\\
    \partial_{\tau} \eta &=  \frac{\mu \sqrt{\mathcal{F}}}{\alpha^2 + \mu^2
    \mathcal{F}} \partial_{\tau} \alpha \; . \label{p_tau_eta}\end{align}
Reinserting \eqref{p_r_eta} and \eqref{p_tau_eta} into \eqref{u_int} or \eqref{u_w2} leads to
\begin{align}\partial_{\tau}\alpha = \frac{\alpha^2}{\mu^2} \partial_R \mu -
  \frac{\mu}{2} \partial_R \mathcal{F} \; .\end{align}
Since $\alpha$ is a function only of $\tau$, and $\mu$ and
$\mathcal{F}$ are functions only of $R$, the above equation requires $\alpha =
\mathrm{const.}$ Hence we have
\begin{align} \frac{\alpha^2}{\mu^3} \partial_{R}\mu = \frac{1}{2} \partial_R
  \mathcal{F} \; ,\end{align}
which yields
\begin{align} \mu = \frac{\beta}{\sqrt{1-a^2\mathcal{F}}} \; ,\end{align}
where $\beta$ and $a= \beta/\alpha$ are constants. Then the unique solutions of
\eqref{W1_old} and \eqref{W2_old} are
\begin{align} W = a \tau + \int \mathrm{d}R
  \frac{\sqrt{1-a^2\mathcal{F}}}{\mathcal{F}} \; , \end{align}
which is just \eqref{uni_factor}.

\end{appendix}


\end{document}